\documentclass[twocolumn,showpacs,preprintnumbers,amsmath,amssymb]{revtex4}

\usepackage{dcolumn}
\usepackage{bm}
\usepackage[pdftex]{graphicx}

\begin{document}

\preprint{APS/123-QED}

\title{Manipulation of Colloids by Nonequilibrium Depletion Force in Temperature Gradient}

\author{Hong-Ren Jiang$^1$}
\author{Hirofumi Wada$^2$}%
\author{Natsuhiko Yoshinaga$^1$}%
\author{Masaki Sano$^1$}
\affiliation{%
$^1$Department of Physics, The University of Tokyo, Hongo 7-3-1, Tokyo 113-0033, Japan \\
$^2$Yukawa Insititute for Theoretical Physics, Kyoto University, Kyoto 606-8502, Japan
}%

\date{\today}

\begin{abstract}
The non-equilibrium distribution of colloids in a polymer solution
under a temperature gradient is studied experimentally. 
A slight increase of local temperature by a focused
laser drives the colloids towards the hot region, resulting in the
trapping of the colloids irrespective of their own thermophoretic
properties. An amplification of the trapped colloid density with
the polymer concentration is measured, and is quantitatively
explained by hydrodynamic theory. The origin of the
 attraction is a migration of colloids driven by a
non-uniform polymer distribution sustained by the polymer's
thermophoresis. These results show how to control
thermophoretic properties of colloids.
\end{abstract}

\maketitle

{\it Introduction.}---
Gradients of thermodynamic variables such as temperature, chemical potential, and osmotic pressure
cause migration of molecules and small particles in fluids\cite{deGroot:1984}. 
For example in biological cell, coupling of two gradients
is often used to promote molecular transport against one of the gradients as in chemiosmosis\cite{mitchell:1978}. 
In physics and chemistry, novel methods to utilize phoretic properties (electro-,  
thermo-, and diffusiophoresis) are proposed for transporting and screening
particles in lab-on-chip\cite{bazant:2004,janca:2003} or designing active materials\cite{Howse:2007}.
In thermophoresis, the speed and direction of migration along a temperature gradient are 
characterized by Soret coefficient, which is generally material-dependent\cite{giglio:1977,de Gans:2003,ning:2006,duhr:2006,braibanti:2008}.
Diffusiophoresis is a similar phenomenon where colloids move along a gradient of 
solute molecules\cite{ebel:1988,anderson:1989,Nardi:1999}. 
The benefits of them have been used for efficient and amplified transport in microfuidics\cite{braun:2002,staffeld:1989,abecassis:2008}. However, strong material dependence of Soret coefficients,
and difficulty in keeping a stationary solute gradient of diffusiophoresis prevent further development
of application of nonequilibrium force by the phoretic effects.
Understanding physical mechanism of thermophoresis has made marked progress recently\cite{degennes:1981,piazza:2004,wurger:2007},
however, it is still challenging to control magnitude of the force to overcome these problems.

 In this paper, we demonstrate that a suitable coupling of thermophoresis for polymer solute molecules
and diffusiophoresis for a target particle (colloid) largely resolve the problems.
More specifically, we report experimental and theoretical studies on a phoretic
motion of colloidal particles in a polymer solution under a temperature gradient.
We find that a Soret coefficient of a colloid  is sensitive to the polymer (PEG) concentration;
as increasing the amount of polymer, the Soret coefficient reverses its sign and its 
magnitude outweighs by far its intrinsic value at the highest polymer
concentration studied. 
The dependence of the Soret coefficient on the polymer is experimentally determined and is corrobolated
by our hydrodynamic calculations.
This novel effect allows us to transport and trap colloids at any desired position
by suitably controlling a temperature distribution and the polymer concentration.

\begin{figure}[b]
\includegraphics[width=8cm]{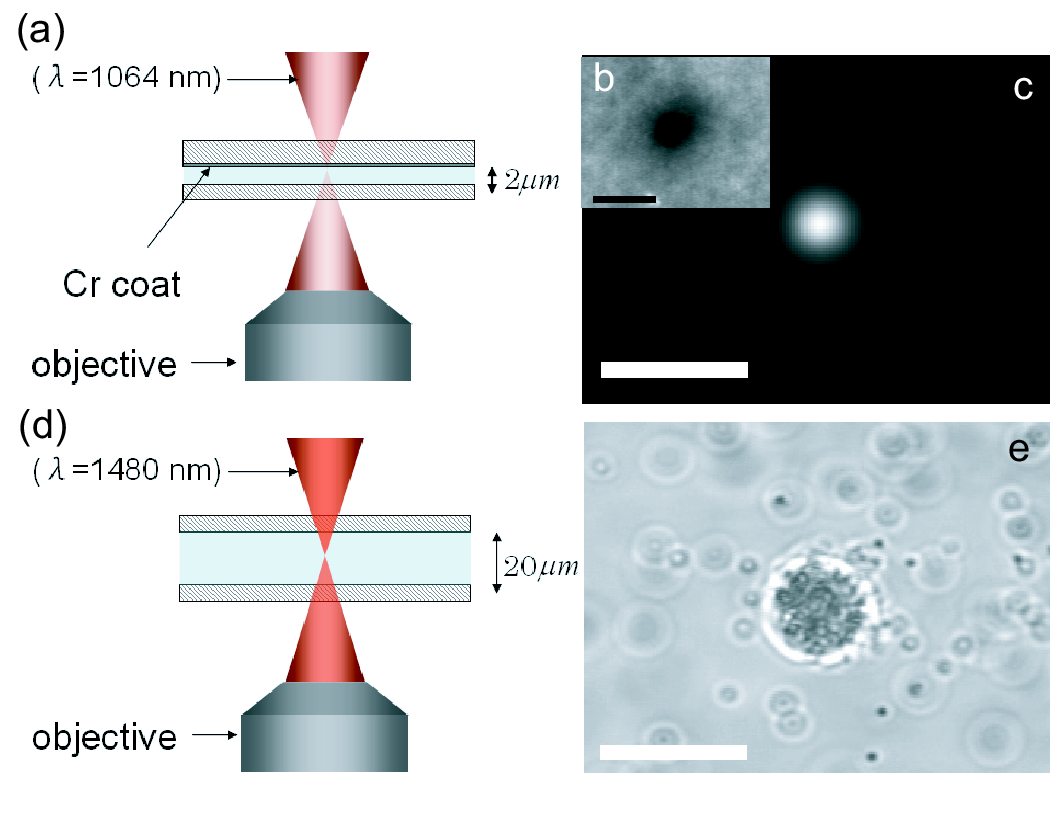}
\caption{ a) (color online) Setup I: The laser fed
 into a thin glass chamber (2$\mu$m) is focused on the top surface
coated by a thin Cr layer for light absorption. b) Distribution of 100 nm fluorescent  polystyrene beads in water. c) Distribution of 100 nm fluorescent polystyrene beads in PEG solution. d) Setup II: The laser fed into a chamber (20$\mu$m thick) is used for directly aqueous solution heating.
e) Cluster of 500nm polystyrene beads in PEG solution. Scale bar: 10$\mu$m}
\end{figure}

{\it Experiment.}--- 
In a thin chamber containing a solution, a steep temperature gradient up
to 1 [K/$\mu$m] was created while keeping the local temperature rise
quite small ($\Delta T\sim4$K) by focusing an infrared laser
(Nd:YAG, 1064 nm, power $\sim$4mW before an objective lens) on a
light absorbing metal coated surface of the thin glass chamber (Setup I) as shown in
Fig.1a \cite{jiang:2007}. Using this setup, we measured the
distributions of polystyrene beads. Figure 1b shows
fluorescence intensity of  100 nm  diameter beads  (F8803; Invitrogen) in a water solution. The fluorescence
intensity became lower at the hot region around laser focus, 
indicating polystyrene beads escaped from the hot region due to thermophoresis. 
The
effects of laser trapping and convection are negligible compared with thermophoresis of
the beads in this setup.
  Polystyrene beads and typical biomolecules such as DNA migrate to colder regions
 at room temperature\cite{duhr:2006}. However, when we added a small
 amount of neutral polymer, polyethylene glycol (PEG6000, MW7500,
 3.5wt\%) in 10mM Tris buffer,  beads and  DNA
 molecules\cite{jiang:2009} migrated and were trapped at the hot region
 regardless of the sign of thermophoresis (Fig.1c). The range of
 attraction reached 5 to 10$\mu$m, corresponding to the size of the
 temperature distribution (see Fig.2a), exceeding by far about 1$\mu$m
 in optical tweezers. As we moved the laser or chamber at several
 micrometers per second, the trapped polystyrene particles moved
 with the hot region (see the supporting information \cite{movie}).
This indicated that the sign and the magnitude of thermophoresis were modified in the presence of polymer under a temperature gradient.
This effect was also observed in a bulk heating configuration as shown in Fig.1d (Setup II). Direct absorption of laser light with longer wavelength(1480nm, 25mW, Furukawa)  in the water produced a local temperature gradient, which creates a 3D colloidal aggregation within the hot region (Fig. 2e)

\begin{figure}[t]
\vspace{-0.30cm}
\includegraphics[width=8.6 cm]{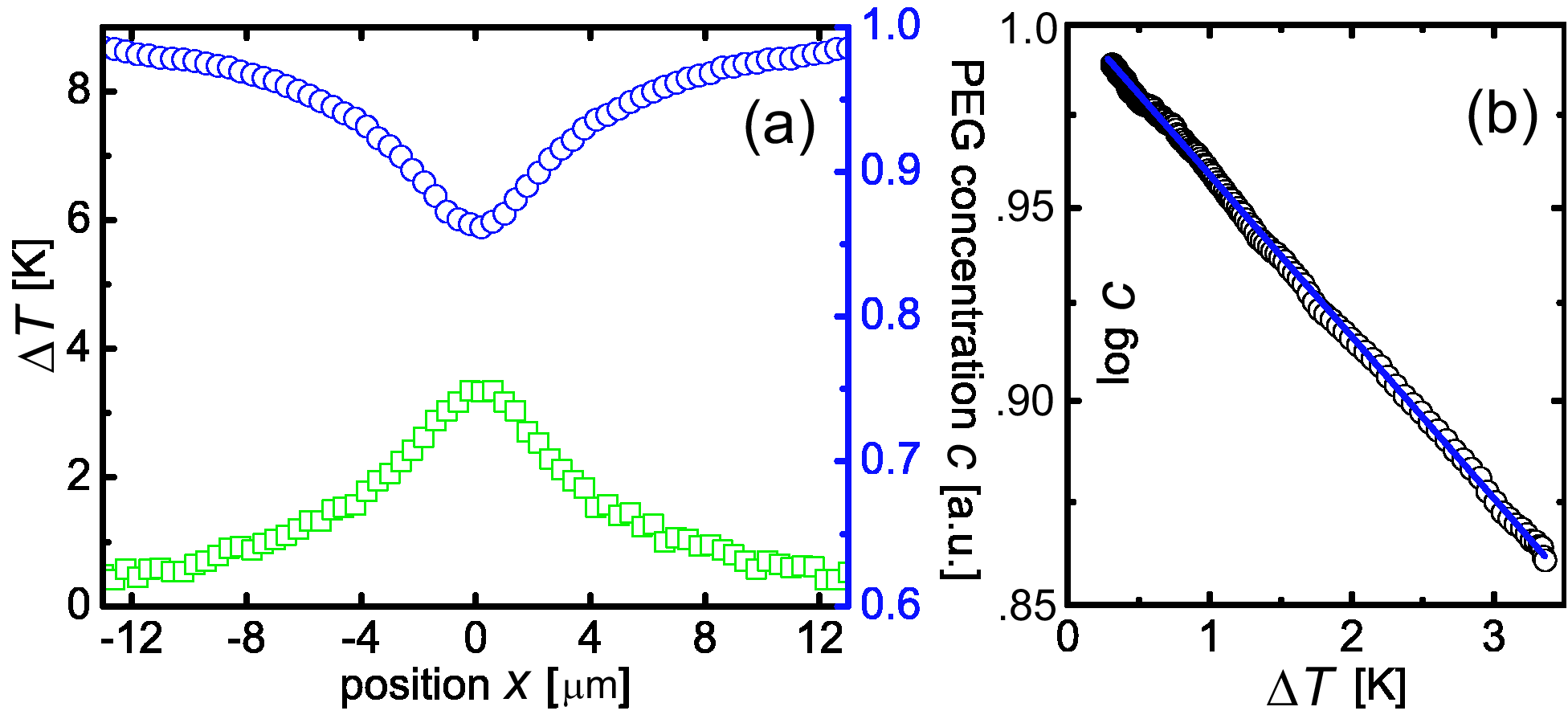}
\caption{\label{fig:epsart2} (color online) a) Profile of temperature (green: rectangle, right ordinate) and density of fluorescent PEG (blue: circle. left ordinate) as a function of radius from the center of laser focus.
b) Density of fluorescent PEG as a function of temperature increment. }
\vspace{-0.30cm}
\end{figure}

\begin{figure*}[t]
\vspace{-0.3cm}
\includegraphics[width=17.5 cm]{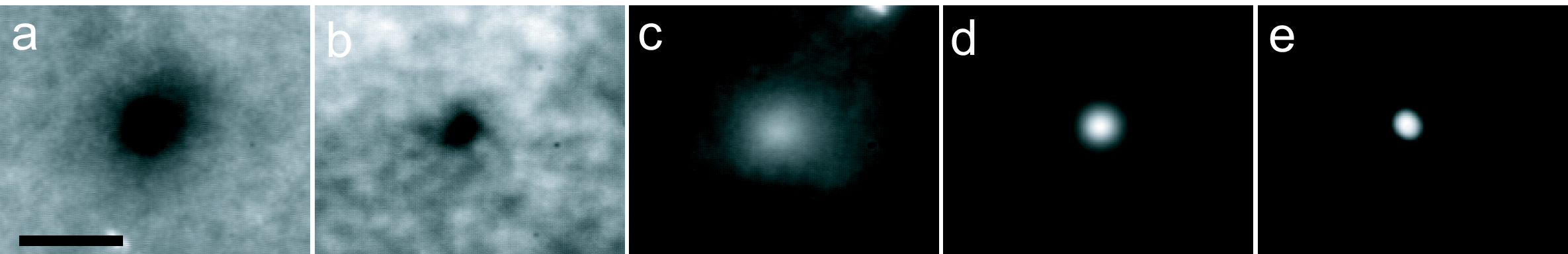}
\caption{\label{fig:wide}Fluorescence intensity of 100 nm beads under temperature gradient for different polymer concentrations,
 a) 0\%, b) 1\%, c) 2\%, d) 3.5\%, and e) 5\% PEG 6000 solutions, respectively. The laser is focused at the center.  The scale bar: 10$\mu$m. }
\vspace{-0.3cm}
\end{figure*}

\begin{figure}
\vspace{-.30cm}
\includegraphics[width=7.5 cm]{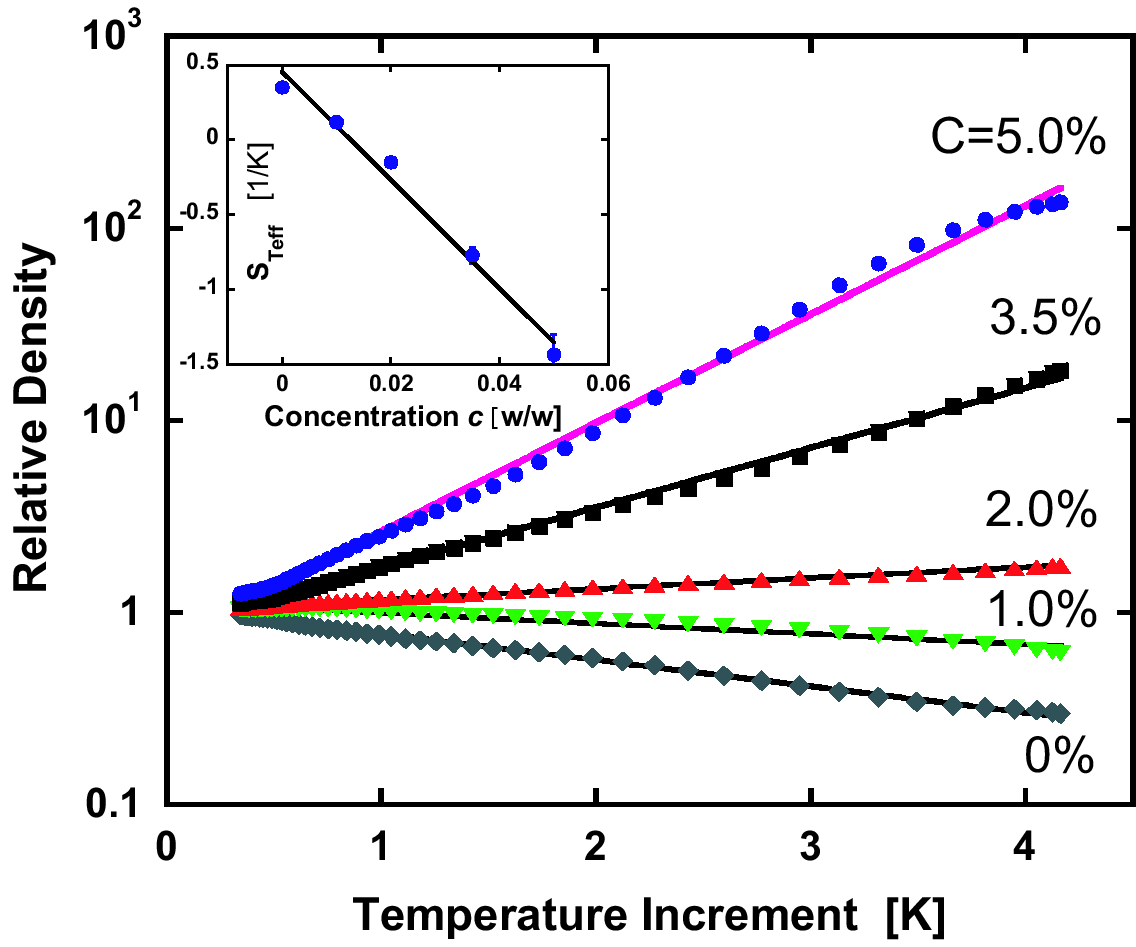}
\caption{\label{fig:Fig4} (color online)  The relationship between the temperature increment and
the density of the beads divided by its background value, for different PEG concentrations indicated.
Filled symbols are the experimental data obtained from Fig.3. The solid lines are the fits with Eq.~(\ref{eq:soret_peg}), from which the effective 
Soret coefficients, $S_T^*$, were obtained. Inset: $S_T^*$ is plotted as a function of the PEG concentration 
in weight fraction, $c$. The solid line is the fitting with Eq.~(\ref{eq:S_T_eff}). }
\vspace{-0.30cm}
\end{figure}

{\it Polymer Distribution.}---
To elucidate the mechanism of the attraction, we measured a profile of temperature and density distribution of PEG near the laser focus in the Setup I. 
The temperature profile and the density of PEG molecules in chamber were shown in Fig.2a\cite{method}. 
 In this setup, temperature profile and density
profile were axisymmetric. Thus the data shown below are averaged along azimuthal direction. The temperature in the laser heating spot increased approximately 1 K with a 1 mW increase in the laser power.  We found that the PEG molecules were depleted by about 15\% due to PEG thermophoresis when the temperature increment was about 3K in the center (Fig. 2a,b). The depletion of PEG can be described as a steady state distribution by the balance between thermodiffusion and diffusion as follows. The flux of PEG molecules obeys the relation, 
\begin{equation}
\bm{j}_p=-D^p \nabla c - c D_T^p \nabla T ,  \label{eq:phoresis}
\end{equation}
where $c, \bm{j}_p, D^p, D_T^p$ are the polymer density, flux, diffusion constant, and thermodiffusion constant, respectively.  In the steady state ($\bm{j}_p=0$), the density satisfies 
\begin{equation}
dc/c = -(D_T^p/D^p)dT = - S_T^{p} dT , \label{eq:soret_diff}
\end{equation}
where the Soret coefficient defined by  $S_T^p = D_T^p/D^p$ is measured by
$S_T^p = -(1/c) (d c/d T) \simeq  -\Delta \ln c/\Delta T$ . 
We obtain $S_T^{p} =0.046 \pm 0.005$ [K$^{-1}$]  for PEG5000 from the slope in Fig.2(b), in agreement with ref.\cite{chan:2004}.

{\it Trapping Ability.}---
Next we examined how the accumulation of beads  is dependent on the concentration of polymer and on the local temperature increase by measuring the density distribution of beads. The fluorescence images of beads were recorded for different PEG6000 concentrations. 
In Fig. 3, spatial distributions of fluorescent polystyrene beads (100 nm, 0.05\%) were displayed.  
Each image was integrated for 300 frames in 20 seconds and
was linearly scaled to 256 grey levels.  In 0\% solution, the density decreased
 from the outer region to the center
(Fig.3a). This depletion of beads is explained quantitatively by thermophoresis,
as described below.  In 1\% solution (Fig. 3b), there was still no
trapping, but the degree of depletion is diminished compared to the 0\% solution. In 2\% solution, the trapping phenomenon began to appear and the fluorescence intensity roughly doubled in the hot region (Fig.3c). In the 3.5\% and 5\% solutions, the density  increased approximately 10- and 100-fold at the center compared with the surrounding concentrations, respectively (Fig.3d,e). 

In Fig.4, the particle density at each distance from the center is plotted against temperature at the same radius (Fig.4). The result for 0\% solution can be described as a steady state distribution due to thermophoresis.  The density of beads, $n$, obeys Eq.~(\ref{eq:soret_diff}) with replacement of $c$ by $n$ and $S_T^{p}$ by $S_T^{b}$, where $S_T^{b}$ is the Soret coefficient of the beads. Neglecting temperature dependence of $S_T^{b}$, an exponential distribution is obtained\cite{braun:2002}
\begin{equation}
n = n_0 \exp[-S_T^{b}(T -T_0)] .  \label{eq:soret_peg}
\end{equation}
 The slope of the curve in
Fig.4 gives the Soret coefficient as $S_T^{b}=0.35\pm0.03$ [K$^{-1}$]  in
agreement with previous results\cite{duhr:2006}. Importantly, even in the presence of polymer, the particle density varied exponentially as a function of the temperature increment (see Fig.4). Therefore a slope of each curves gives an effective Soret coefficient, $S_T^{*}$. We plot $S_T^{*}$ as a function of PEG concentration in the inset of Fig.4. As a first-order approximation, $S_T^{*}$ decreased linearly with the increase of polymer concentration according to the relation
\begin{equation}
S_T^{\ast}= S_T^{b} - \chi c
\label{eq:S_T_colloid}
\end{equation}
with $\chi= 35.4 \pm 2.0$ [K$^{-1}$] as the best fit, and where $c$ is the concentration of PEG in weight fraction, or $\chi= 441\pm 25$[nm$^{3}$K$^{-1}$] for $c$ in volume density. Hence, the sign and the magnitude of the Soret coefficient was controlled by varying the concentration of polymer. 

\begin{figure}[t]
\includegraphics[width=8 cm]{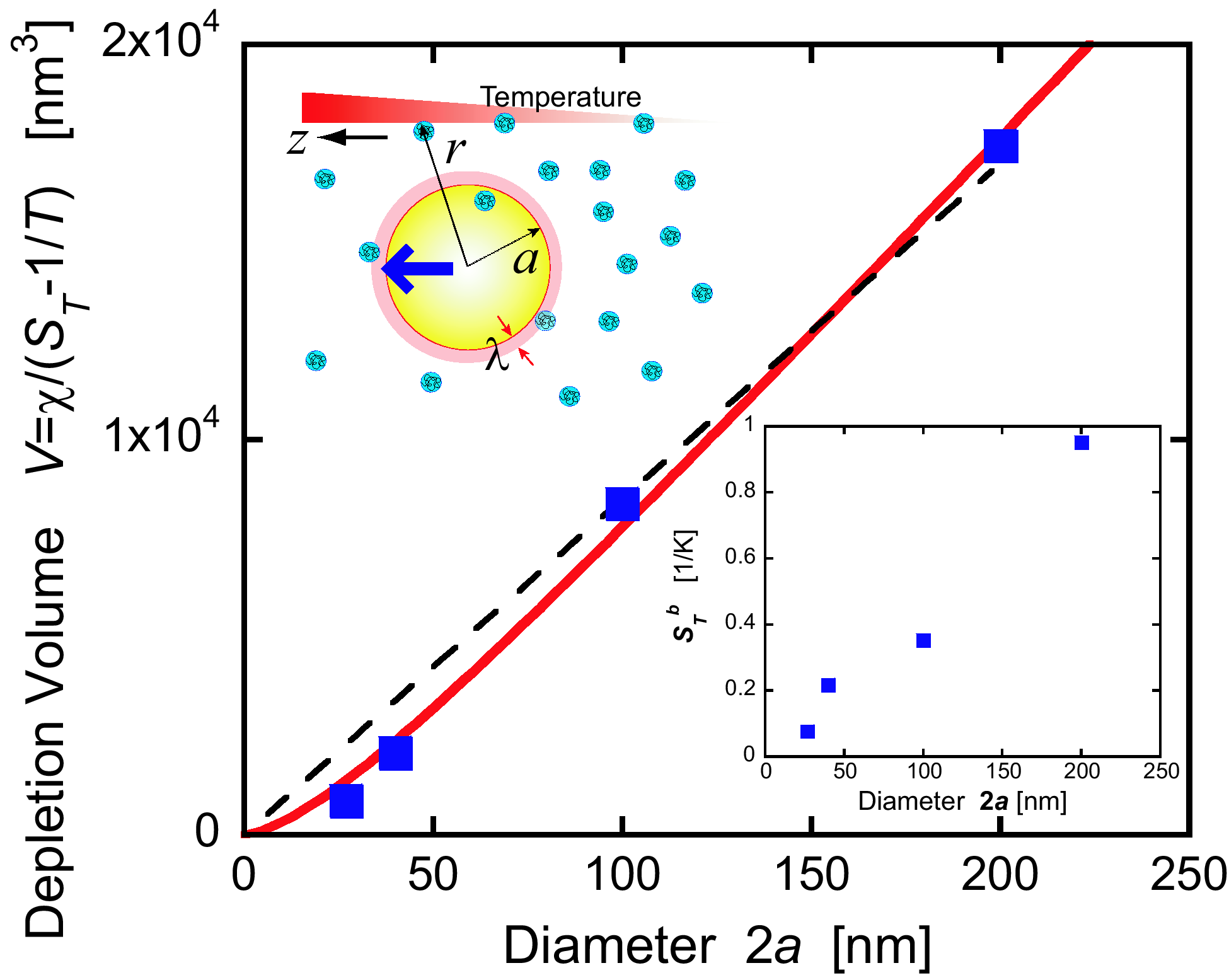}
\caption{\label{fig:Fig5} (color online) Particle diameter dependence of the depletion volume $V$. Dotted line and solid line represent
theoretical predictions for non-slip and Navier boundary condition, respectively. Inset : (top) schematic illustration of a colloid particle in polymers solutuion. (bottom) Soret coefficients of beads for different sizes.}
\vspace{-0.3cm}
\end{figure}

{\it Theory.}---
The main results obtained in our experiment are quantitatively explained
by the hydrodynamic
theories\cite{anderson:1989,piazza:2004,wurger:2007}.
We consider a spherical particle of radius $a$ in a
polymer solution, see the inset of Fig.~\ref{fig:Fig5}.
In a dilute regime valid to our experiments, PEG distribution
around the particle obeys Boltzmann distribution,
$c = c_0 \exp(-U(r)/k_B T)$, where $c_0$ is the concentration at infinity, $k_B$ is the Boltzmann constant, 
and $U(r)$ is a short range potential for PEG molecules.
As a PEG is non-ionic and inactive, $U$ represents an entropic
repulsion from the colloid surface with its interaction distance $\lambda$.
Note that $\lambda$ also defines a length scale of the depletion interaction between
colloids at equilibrium\cite{asakura:1958,croker:1999}.
In a temperature gradient $dT/dz$ applied along $z$ direction, a PEG
gradient $dc_0/dz$ is developed according to Eq.~(\ref{eq:soret_peg}).
As $c_0$ changes only gradually at a scale of  $\lambda$, the steady state distribution is given by 
$c \simeq c_0(z) e^{-U(r)/k_B T}$. Neglecting interactions between polymers,
 an osmotic force density is given by $\bm{f}=-c(\bm{r}) \nabla U(\bm{r})$, 
which is written in, 
$\bm{f} = \nabla (k_B T \delta c) + k_B T (S_T^{p} - \frac{1}{T}) \delta c \nabla T$,
where $\delta c \equiv c - c_0= c_0 (e^{-U/k_B T} -1)$, 
and $\nabla c_0 = - c_0 S_T^{p} \nabla T$ was used.
The velocity field around the colloidal sphere, $\bm{v}$, is obtained by
solving the Stokes equation,
$\eta \nabla^2 \bm{v} = \nabla p - \bm{f} $, with the incompressibility
condition, $\nabla \cdot \bm{v}= 0$,
where $\eta$ is the viscosity of the solution and $p$ is the hydrostatic
pressure.
The Stokes equation is rewritten in the following transparent form,
\begin{equation}
\eta \nabla^2 \bm{v} = \nabla (p-k_B T \delta c) + f_0(r)\bm{e}_z ,
\label{eq:stokes}
\end{equation}
where  $f_0 (r) \equiv k_B T \bar{c} (S_T^p - 1/T) (1 - e^{-U/k_BT}) dT/dz$ and $\bar{c}=c_0(z=0)$. 
In Eq.~(\ref{eq:stokes}), the osmotic pressure,  $k_BT\delta c$, is compensated by a hydrostatic pressure $p$.
(This ensures the absence of osmotic force on the particle proportional to its volume, $\sim k_BTa^3\delta c$.) 
The nonequilibrium force, $f_0\bm{e}_z$ in Eq.~(\ref{eq:stokes}), on the other hand, is balanced with the shear viscous force, 
$\eta\nabla^2\bm{v}$, leading to a fluid flow relative to the colloid surface and 
a phoretic motion of the colloid at a velocity $\bm{u}$ relative to the fluid at infinity.
Solving Eq.~(\ref{eq:stokes}) in the colloid-reference frame with the no-slip boundary
condition,
and summing up the resulting fluid stress over the colloidal surface, we
obtain the total force acting on the colloid as
${\bf F} = - 6 \pi \eta a \bm{u} + 4 \pi \bm{e}_z \int_{a}^{\infty} r (r-a)  f_0
dr$\cite{piazza:2004}.
This must be zero because no external force is acting on the colloid.
Choosing a specific potential of hard-core type (i.e.,
$U=\infty$ for $a<r<a+\lambda$ and zero otherwise), we finally arrive at
\begin{equation}
 {\bf u}= \frac{k_B T}{3 \eta}(S_T^{p}-\frac{1}{T}) \lambda^2\bar{c}
\nabla T.
 \label{eq:velocity}
\end{equation}
Plugging then Eq.~(\ref{eq:velocity}) into an expression of the density
current of colloids,
$\bm{J} = n\bm{u} - D \nabla n - n D_T \nabla T$\cite{wurger:2007}, and rewriting it in a
form $\bm{J}=- D \nabla n - n D S_T^{\ast}  \nabla T$,
we see that an effective Soret coefficient is given by
\begin{equation}
S_T^{*} = S_T^{b} - 2 \pi (S_T^{p} -\frac{1}{T})  a \lambda^2 \bar{c},
\label{eq:S_T_eff}
\end{equation}
which confirms the linear dependence on $c$ in Eq.~(\ref{eq:S_T_colloid}), 
where $\chi=2\pi(S_T^p-1/T)a\lambda^2$. 

Equation~(\ref{eq:S_T_eff}) is compared to our experimental data in Fig.~\ref{fig:Fig4} inset by employing
$S_T^p=0.056$ for  PEG with MW=7500 estimeted from \cite{chan:2004}. 
The best fit gives $\lambda \simeq 5.2$ nm, which is close to the PEG gyration radius
$\sim 3$ nm~\cite{kawaguchi:1997}. 
We consider the agreement satisfactory, as $\lambda$ should be of the order of a PEG size.
A further consistency check was made by looking at the dependence of $S^{\ast}_T$ on the particle radius $a$.
To focus only on this dependence of $S_T^{\ast}$,  
we define an effective depletion volume at nonequilibrium
$V=\chi/(S_T^p-1/T)$.  
First, we performed the series of experiments for four different colloid sizes, and extracted $V$ from the data on $S_T^{\ast}$,
which are plotted in Fig.~\ref{fig:Fig4} as a function of the diameter $2a$.
The data shows an overall linear dependence, and the agreement with the theoretical prediction for
no-slip condition, $V=2\pi \lambda^2 a$, obtained from Eq.~(\ref{eq:S_T_eff}) is good. 
Note that we used $\lambda=5.2$ nm obtained before, and thus no adjustable parameter is assumed.
Nevertheless, it would be important to examine effects of slip of the polymer solution upon the enhancement of
the Soret effect, since the perfect slip condition predicts a different scaling, $V\sim \lambda a^2$~\cite{wurger:2007}.
A boundary condition for a general slip case is a so-called Navier
condition~\cite{happel:1965}, given by
$\sigma_{t}=\eta v_{t}/\ell$ at $r=a$, where $\sigma_{t}$ and $v_{t}$ are
the tangential components of the fluid stress and velocity.
The parameter $\ell$ defines a slip length on the surface (the no-slip or
perfect slip limit is attained for $\ell=0$ or $\ell=\infty$)\cite{ajdari:2006}.
In this Navier case we obtain 
$V\simeq 2\pi a\lambda [a\lambda+2\ell(a+2\lambda)]/(a+3\ell)$.
This formula, with $\lambda=1.5$ nm and $\ell=9.7$ nm as the best
fit, actually improves the agreement with the data
for $a<50$ nm, where the deviation from the linear dependence is
observable (see Fig.~\ref{fig:Fig5}).
In our system, therefore, a fluid slip might be relevant for $a<50$ nm (with a relatively large
slip lengh $\ell\sim 10$ nm).

{\it Summary.}---
We have developed a micro-manupulation technique allowing us to invert
and amplify the movement of colloid particles due to thermophoresis of
polymers induced by laser focusing. The polymer concentration
 gradient created by thermophoresis causes migration of 
a particle with a speed determined by the balance between the driving force and the viscous force.
This new method and its application to trapping of molecules based on
nonequilibrium effects does not rely on 
a specific character of particles and polymers, and
 thus provides further applications for manipulating a diverse range of colloidal particles as well as
biological cells and DNA molecules\cite{jiang:2009,ichikawa:2005}.
Our technique is also applicable with other
water-soluble molecules\footnote{We confirmed
that  polyvinylpyrrolidone
(PVP) and sodium polystyrene sulfonate (NaPSS) also work, yet the effect is strongest
for PEG.}. We also note that the effect is stronger for PEG with a larger molecular weight
at the same monomer concentration as far as it is lower than the overlap concentration.

This works is supported by Grant-in-Aid for Scientific Research from MEXT Japan.

\end{document}